\documentclass[eqsecnum,floats,aps,nofootinbib,showpacs]{revtex4}
\usepackage{latexsym}
\usepackage{amssymb}

\usepackage[pdftex]{hyperref}

\renewcommand{\d}{\mathrm{d}}
\renewcommand{\l}{\left(}
\renewcommand{\r}{\right)}
\usepackage{amsmath, amsthm, amssymb}

\def\be{\begin{equation}}
\def\ee{\end{equation}}
\def\beq{\begin{equation*}}
\def\eeq{\end{equation*}}
\def\ba{\begin{aligned}}
\def\ea{\end{aligned}}

\def\ov{\overline}
\def\w{\wedge}

\begin{document}
\title{Translations in Quantum Field Theory and the Poincar{\'e} Gauge Theory of Gravity}

\author {Marcin Ka\'zmierczak}
\email{marcin.kazmierczak@fuw.edu.pl}
\affiliation{Institute of Theoretical Physics, Uniwersytet Warszawski, Ho\.{z}a 69, 00-681 Warszawa, Poland} 

\begin{abstract}
In standard quantum field theory, the one--particle states are classified by the unitary representations of the Poincar{\'e} group, whereas the causal fields' classification employs the finite dimensional (non--unitary) representations of the (homogeneous) Lorentz group. We investigate the possibility of constructing fields that transform under the full representation of the Poincar{\'e} group. We show that such fields can be consistently constructed, although the Lagrangians that describe them exhibit explicit dependence on the space--time coordinates. The inclusion of gravity within the framework of the Poincar{\'e} gauge theory is then discussed. A new feature that occurs is that the translational gauge fields enter the covariant derivative of matter fields. The Poincar{\'e}--gauge approach to gravity works still well and leads to interesting consequences. The detailed discussion of the Dirac field is presented and the relation to the earlier accounts on Poincar{\'e}--spinors is drawn. Another example that is considered is the Poincar{\'e}--vector field. The presentation has a partly didactic character and is addressed to all the readers who are interested in the rudiments of quantum field theory and the gauge description of gravity.  
\end{abstract}
\maketitle

\section{Introduction}

Assume that gravity is sufficiently small to be neglected in some sort of experiments. Then special relativity implies that every theory accounting for the results of these experiments need to be invariant under the global action of the Poincar{\'e} group corresponding to the passage from one inertial observer to another. The basic idea of the Poincar{\'e} gauge theory of gravity (PGT) is to introduce gravity by localization of this fundamental symmetry. Since the pioneering work of Yang and Mills on the strong interactions \cite{YaMi}, such an approach to the description of interactions that is based on localization of global symmetries has proved extremely successful in the non--gravitational sector. The idea of describing gravity in a similar way has now a long history. Initially, gravity was viewed as a gauge theory of the Lorentz group by Utijama \cite{Ut}. Then Kibble \cite{Kib1} observed that promoting the whole Poincar{\'e} group to the gauge group has a lot of  advantages. Among them, one  is not forced to introduce a cotetrad on space--time {\it ad hoc} -- it can be related to the translational gauge fields and derived via the localization procedure, in much the same way as a space--time connection. Further investigations of the idea where made (see e.g. \cite{HHKN} for a review), but the translational and homogeneous parts of the group were not really treated on the same footing until the work of Grignani and Nardelli \cite{GN}, where the authors realized that 
only after additional fields are introduced on space--time can the theory be cast into the form that is truly similar to the geometric setting of standard Yang--Mills theories. These fields, called the {\it Poincar{\'e} coordinates}, transform as Poincar{\'e} vectors under gauge transformations. In fact, their geometric interpretation in terms of the theory of connections on a principal fiber bundle of affine frames was given much earlier by Trautmann \cite{T1}, but the physical interpretation in the usual formulation of the theory in terms of sections of the bundle was not discussed there. What is more, the complete affine group was considered as a gauge group, rather than merely its Poincar{\'e} subgroup. For an exhaustive review of possible approaches to the gauge formulation of gravity, see \cite{HCMN}. It is important to remember that the Poincar{\'e} group remains the most natural candidate for the gauge group from the physical point of view because of its relation to the principle of equivalence of special relativity. We shall elaborate on the Poincar{\'e} gauge theory, and in particular on the inclusion of matter in such a theory, in Section \ref{section2}. 

In Section \ref{QFT} we shall restrict ourselves to the circumstances of negligible gravity and explore the structure of global Poincar{\'e} invariance.
In conventional quantum field theory (QFT), the Hilbert spaces of one--particle states are constructed by considering irreducible unitary representations of the Poincar{\'e} group\footnote{More precisely, its universal covering group, if spinor representations are to be included. This remark should be understood to hold throughout the paper.}
derived by induction from irreducible representations of the little group which are labeled by spin (massive case) or helicity (massless case). Then the quantum causal fields are introduced that are classified by finite--dimensional representations of the (homogeneous) Lorentz group. Any such representation, acting on the space of fields,  needs to be connected to some unitary representation of the Poincar{\'e} group, acting on the Fock space of many--particle states, via Weinberg consistency conditions. These conditions are in fact equations for the amplitudes of the fields. For a fixed representation of the Lorenz group, the consistency conditions restrict the set of allowable representations of the Poincar{\'e} group on the Fock space. 
Choosing one of these representations and solving the conditions leads to the form of the amplitudes that correspond to a unique  value of spin and can be interpreted as describing a particular kind of particles. Further restrictions on the amplitudes follow from causality conditions, charge conservation, self--adjointness of the interaction density and possibly discrete symmetries. The fields thus constructed obey Lorentz invariant equations which can be given Lagrangian formulation. Then, the conserved currents of physical interest can be inspected by means of Noether theorem, the interaction terms and the $S$--matrix can be constructed. See the classical reference \cite{Wein} for a review of this approach to QFT.

In fact, the Lagrangians of such theories are invariant under the global action of the Poincar{\'e} group where the translations act trivially. The Lorentz group, on the other hand, can act trivially (e.g. for a scalar field) or nontrivially (e.g. for a vector or a spinor field). The latter possibility has the consequences for PGT, as it leads to the presence of Lorentz gauge fields in the covariant derivative of the field after gravity is included. These gauge fields give rise to the connection on the space--time manifold. The situation is much different for translational gauge fields. Since the representations of the Poincar{\'e} group acting on the space of fields are not faithful (all momentum generators are represented by zero operators), the translational gauge fields do not enter the covariant derivatives of matter fields constructed according to the standard minimal coupling procedure (MCP). Although this state of affairs does not contradict the consistency or physical relevance of PGT, it seems somewhat unsatisfactory. The situation can be compared to the one in which we knew that 
electromagnetism, say, ought to be viewed as a gauge theory of the $U(1)$ group, but all the field theories that we were aware of contained real fields only. Certainly, the analogy does not go too far, since the translational gauge fields enter the theory anyway due to their relation to the Poincar{\'e} coordinates (although they can then appear as hidden in the cotetrad only -- see Section \ref{section2}). However, the issue of nontrivial implementation of translations in PGT and, in particular, the question of how the fermionic matter should be included in such a theory attracted a lot of interest in the course of time (see e.g. \cite{GN}\cite{Lec1}\cite{TT1}\cite{TT2}).

 A natural question is in order:  what is the reason for classifying causal fields by representations of the Lorentz group only and not the full Poincar{\'e} group? In other words, why are all the finite--dimensional representations of the Poincar{\'e} group that classify the fields non--faithful? We will try to answer this question in Section \ref{QFT}. We shall show that the fields that transform under the faithful finite--dimensional representations of the Poincar{\'e} group can be consistently constructed within the framework of QFT. We shall investigate their properties and clarify the relation to conventional fields that transform trivially under translations. The inclusion of gravity is discussed in Section \ref{section2}. In section \ref{conc} we draw the conclusions.

\section{Quantum field theory with nontrivial translations}\label{QFT}

\subsection{General formalism}

In this section we shall parallel the basic derivations of \cite{Wein} allowing for the presence of nontrivial generators of momenta in the representations that classify the 
quantum fields.
Let $U(\Lambda,b)=U(b)U(\Lambda)$, $U(b)=\exp\l{ib\cdot P}\r, \ U\l{\Lambda(\varepsilon)}\r=\exp\l{\frac{i}{2}\varepsilon_{ab}J^{ab}}\r$ be the irreducible unitary representation of the universal covering of the Poincar{\'e} group. Here $P^a, J^{ab}$ are the self--adjoint generators of translations and Lorentz rotations belonging to the corresponding representation of the Poincar{\'e} algebra, $\varepsilon\equiv\l{{\varepsilon^a}_b}\r\in so(1,3)$ is such that $\Lambda\l{\varepsilon}\r{^a}_b={\delta^a}_b+{\varepsilon^a}_b+\dots$, $b\equiv\l{b^a}\r$ are the parameters of space--time translation and $\cdot$ denotes the Minkowski product, $b\cdot P=\eta_{ab}b^aP^b$, $\eta=diag(1,-1,-1,-1)$. We shall restrict ourselves to the massive representations for which $P\cdot P=m^2$, where $m>0$ is interpreted as a mass of a particle. Then the little group that leaves the standard momentum $k=(m,0,0,0)$ invariant is $SO(3)$ and its universal covering is $SU(2)$. Let $D^j$ denote the irreducible unitary representations of $SU(2)$ labeled by spin $j=0,\frac{1}{2},1\dots$\footnote{Note that the expressions of the form $D^j(R)$ for $R\in SO(3)$, which we shall use for simplicity, are in general multi--valued.}. Let $\Psi_{p,\sigma}$ be the (distributional) basis of the space of one--particle states with well established value of momentum $p$ and the projection of spin on the third spatial axis in the rest frame $\sigma$. The action of the representation $U$ on this basis is
\be\label{Utr}
U(\Lambda,b)\Psi_{p,\sigma}=e^{ib\cdot\Lambda
  p}D^j\l{W_{\Lambda,p}}\r_{\sigma'\sigma}
\Psi_{\Lambda p,\sigma'} \quad ,
\ee
where
$L_p\in SO(1,3)$ is the standard bust, $L_pk=p$, given explicitly by 
\be\label{bust}
L_p=
\l
\begin{array}{cccc}
\frac{p^0}{m}\qquad & \frac{\mathbf{p}^T}{m} \\
\frac{\mathbf{p}}{m}\qquad & {\mathbf 1}_3+\frac{\mathbf{p}\mathbf{p}^T}{m(p^0+m)} 
\end{array}
\r
\ee
(think of $\mathbf{p}\in\mathbb{R}^3$ as a column matrix)
and 
$W_{\Lambda,p}={L_{\Lambda p}}^{-1}\Lambda L_p$ belongs to the little group of $k$. We shall follow the normalization convention according to which
$\Psi_{p,\sigma}=U(L_p)\Psi_{k,\sigma}$.
The one--particle states are created from vacuum according to $a^{\dag}_{p,\sigma}\Psi_0=\Psi_{p,\sigma}$,
where $\Psi_0$ represents the vacuum state, normalized to unity. The commutation relations for creation and annihilation operators are adopted in the form 
\be
[a_{p,\sigma},a^{\dag}{}_{p',\sigma'}]_{\mp}=(2\pi)^32p^0\delta({\bf
  p}-{\bf p}')\delta_{\sigma,\sigma'}\quad,
\quad [a_{p,\sigma},a_{p',\sigma'}]_{\mp}=[a^{\dag}_{p,\sigma},a^{\dag}_{p',\sigma'}]_{\mp}=0\quad,
\ee
where the upper sign in $\mp$ denotes the commutator and refers to bosons, while the lower one denotes the
anti--commutator and refers to fermions (in all the formulas it should be understood that we are on the mass shall, i.e. $p^0=\sqrt{{\bf p}^2+m^2}$). Assuming the invariance of the vacuum $U(\Lambda,b)\Psi_0=\Psi_0$, one can derive from (\ref{Utr}) the transformation law for the creation operators
\be\label{atr}
U(\Lambda,b)a^{\dag}_{p,\sigma}U^{-1}(\Lambda,b)=e^{ib\cdot\Lambda
p}D^j\l{W_{\Lambda,p}}\r_{\sigma'\sigma}a^{\dag}_{\Lambda p,\sigma'}\quad,
\ee
from which the transformation law for the annihilation operators follows by conjugation.

Let us introduce creation and annihilation fields 
\be
\psi^{+}_l(x)=\int \, u_{l\sigma}(x,p)a_{p,\sigma} \, d\Gamma_p\quad , \quad
\psi^{-}_l(x)=\int \, v_{l\sigma}(x,p)a^{\dag}{}_{p,\sigma}\, d\Gamma_p \quad ,
\ee
where $d\Gamma_p=\frac{d^3p}{(2\pi)^32p^0}$ is the Lorentz--invariant measure.
We shall require that these fields satisfy the transformation law
\be\label{psitr}
U(\Lambda,b)\psi^{\pm}_l(x)U^{-1}(\Lambda,b)=\rho^{-1}_{ll'}(\Lambda,b)\psi^{\pm}_{l'}(\Lambda
x+b) \quad ,
\ee
where 
\be\label{Poinrep}
\rho(\Lambda,b)=\rho(b)\rho(\Lambda)\quad,\quad \rho(b):=\rho\l{{\bf 1},b}\r=\exp\l{ib\cdot\mathbb{P}}\r\quad,\quad 
\rho(\Lambda):=\rho\l{\Lambda(\varepsilon),0}\r=\exp\l{\frac{i}{2}\varepsilon_{ab}\mathbb{J}^{ab}}\r
\ee
is a finite--dimensional (non--unitary) representation of the Poincar{\'e} group. Thus the formula differs from the one considered in \cite{Wein} by the presence of the non--zero momentum generators $\mathbb{P}^{a}$. Note that for $\Lambda={\bf 1}$ and infinitesimal $b$ (\ref{psitr}) implies that
\be
[P_a,\psi^{\pm}(x)]_{-}=\l{-i\partial_a-\mathbb{P}_a}\r\psi^{\pm}(x)\quad .
\ee

Using (\ref{atr}), the consistency conditions relating the representations $U$ and $\rho$ can be derived
\be\label{war}
\ba
\rho^{-1}(\Lambda,b) u(\Lambda x+b,\Lambda p)&=e^{-ib\cdot\Lambda p} u(x,p){D^j}^{-1}\l{W_{\Lambda,p}}\r \quad , \\ 
\rho^{-1}(\Lambda,b) v(\Lambda x+b,\Lambda p)&=e^{ib\cdot\Lambda p} v(x,p){D^j}^T\l{W_{\Lambda,p}}\r \quad ,
\ea
\ee
where $u$ and $v$ denote matrices whose entries are $u_{l\sigma}$ and $v_{l\sigma}$) respectively and $^T$ stands for the transposition of a matrix. 
For pure translations one gets 
\be
u(x+b,p)=e^{-ib\cdot p}\rho(b)u(x,p) \quad , \quad v(x+b,p)=e^{ib\cdot p}\rho(b)v(x,p) \quad .
\ee
The solution is provided by the following form of the amplitudes $u$ and $v$
\be\label{uxvx}
u(x,p)=e^{-ip\cdot x}\rho(x)u(p) \quad, \quad v(x,p)=e^{ip\cdot
  x}\rho(x)v(p) \quad,
\ee
where $u(p)\equiv u(0,p)$ and $v(p)\equiv v(0,p)$ do not depend on $x$. 
Thus the fields $\psi^{\pm}(x)$ are not just the Fourier transforms, as in the conventional QFT.  Inserting (\ref{uxvx}) into (\ref{war}) and
employing the composition law
$\rho(\Lambda,a)\rho(\Lambda',a')=\rho(\Lambda\Lambda',\Lambda a'+a)$ we get the standard Weinberg conditions
\be
u(\Lambda p)=\rho(\Lambda)u(p){D^j}^{-1}\l{W_{\Lambda,p}}\r \quad, \quad
v(\Lambda p)=\rho(\Lambda)v(p){D^j}^{T}\l{W_{\Lambda,p}}\r \quad.
\ee
Hence, the $x$--independent parts of the amplitudes satisfy the conditions of standard theory. In particular, it follows that
\be\label{pchobr}
u(p)=\rho(L_p)u(k) , \quad v(p)=\rho(L_p)v(k) , \quad
u(k)=\rho(R)u(k){D^j}^{-1}(R) , \quad v(k)=\rho(R)v(k){D^j}^T(R) ,
\ee
for the standard momentum $k$ and any rotation $R$. Only the dependence of $\psi^{\pm}$ on $x$ is changed by the presence of $\rho(x)$. 
The last two equations of (\ref{pchobr}) tell us simply that $u(k)$ plays a role of a morphism between the representations $\rho(R)$ and $D^j(R)$, whereas $v(k)$ is a morphism between $\rho(R)$ and $D^{j*}(R)$. The representations $D^j$ (and $D^{j*}$) are irreducible. If $\rho$ provided an irreducible representation as well when restricted to the rotational subgroup of the Poincar{\'e} group, then Shur's lemma would imply that the amplitudes either vanish or are isomorphisms (i.e. square matrices). In general, however, $\rho(R)$ is not irreducible, but rather acquires (in the appropriate basis) a block--diagonal form
\be
\rho(R)=
\left({
\begin{array}{cccc}
\rho^1(R) & 0 & 0 \\
0 & \ddots & 0 \\
0 & 0 & \rho^M(R)
\end{array}
}\right)\quad ,
\ee
where the representations $\rho^i$ are irreducible and zeros mean zero matrixes of appropriate shapes and dimensions. If the amplitudes are divided correspondingly as
\be\label{uivi}
u(k)=\left({
\begin{array}{ccc}
u^1 \\
\vdots \\
u^M
\end{array}
}\right)\quad,\quad
v(k)=\left({
\begin{array}{ccc}
v^1 \\
\vdots \\
v^M
\end{array}
}\right)\quad,
\ee
where the number of rows of the matrices $u^i$, $v^i$ equals the dimension of the corresponding representation $\rho^i$, then the last two equations of(\ref{pchobr}) will reduce to the collection of matrix equations $\rho^i(R)u^i=u^iD^j(R)$, $\rho^i(R)v^i=v^iD^{j*}(R)$, $i=1,\dots M$. All the representations that occur here are now irreducible and hence Shur's lemma applies. For a fixed value of spin $j$ it follows that $u^i$ (or $v^i$) can be nonzero only if the corresponding representation $\rho^i$ has dimension $2j+1=dim\l{D^j}\r=dim\l{D^{j*}}\r$. Hence, the representation $\rho$ can describe a particle with spin $j$ only if it contains at least one $2j+1$--dimensional irreducible representation of the group of rotations. After the last two equations of (\ref{pchobr}) are solved, the amplitudes are composed of blocks of $(2j+1)\times (2j+1)$--dimensional non--zero matrices, possibly separated by some blocks of zeros. This zeros may seem superfluous at first, but they may be filled by non--zero expressions when the amplitudes are busted (e.g. for Lorentz--vector field of spin $j=1$). Also, the different non--zero blocks that transform completely independently under rotations may be mixed by discrete symmetries such as parity (e.g. the Dirac field) or by the action of the complete representation $\rho$ of the Poincar{\'e} group (e.g. Poincar{\'e}--vector field, see Section \ref{Poinv}).

In order to satisfy requirements of conservation of electric charge and self--adjointness of an interaction density composed of causal fields, it is necessary to consider the combinations \cite{Wein}
\be
\psi(x)=\psi^{+}(x)+{\psi^{-}}^c(x)\quad, 
\ee 
where ${\psi^{-}}^c(x)=\int d\Gamma_p v(x,p){a^c}^{\dag}_{p}$ and $c$ stands for anti--particle (for electrically neutral particles $a^c=a$).
The fields $\psi^{\pm}$ and ${\psi^{\pm}}^c$, and hence also $\psi$, 
are assumed to transform according to the same representation $\rho$ of the Poincar{\'e} group.
The fields should also satisfy the causality condition -- the commutator
\be\label{przy}
\ba
&[\psi_l(x),\psi^{\dag}_{l'}(x')]_{\mp}=\left[{\rho(x)\int\l{
e^{-ip\cdot (x-x')}N(p)\mp
e^{ip\cdot (x-x')}M(p)}\r d\Gamma_p \ \rho^{\dag}(x')}\right]_{ll'}, \\
&N(p)=u(p)u^{\dag}(p), \quad M(p)={v^c(p)v^c}^{\dag}(p),
\ea
\ee
ought to vanish for space--like interval $x-x'$ (also
$[\psi_l(x),\psi_{l'}(x')]_{\mp}$ should satisfy the condition, but 
for charged particles it is fulfilled automatically). 

\vskip 0.1 in
\noindent{\bf Parity transformation}

If there are many non--zero blocks $u^i$ ($v^i$) in the decomposition (\ref{uivi}) of the amplitudes, then the relative weights of these blocks will not be fixed by equations (\ref{pchobr}). One can make use of parity transformation to limit this arbitrariness. Let $\mathcal{P}=diag(1,-1,-1,-1)$ represent the parity operation in Minkowski space. In order for the quantum theory to be parity invariant, there should exist a unitary transformation ${\mathrm P}$ acting on the space of states and satisfying the following commutation relations with the generators of the unitary representation $U$ of the Poincar{\'e} group
\be\label{parcom}
\mathrm{P}P^a\mathrm{P}^{-1}={\mathcal{P}_b}^aP^b\quad,
\quad \mathrm{P}J^{ab}\mathrm{P}^{-1}={\mathcal{P}_c}^a{\mathcal{P}_d}^bJ^{cd}\quad.
\ee
The action of this transformation on annihilation and creation operators is
\be
\mathrm{P}a_{p,\sigma}\mathrm{P}^{-1}=\eta^*a_{\mathcal{P}p,\sigma},\quad
\mathrm{P}{a^c}^{\dag}_{p,\sigma}\mathrm{P}^{-1}=\eta^c{a^c}^{\dag}_{\mathcal{P}p,\sigma}\quad,
\ee
where $\eta$ ($\eta^c$) is the internal parity of the particle (anti--particle). The action of parity on fields is then 
\be\label{parpsi}
\ba
&\mathrm{P}\psi^+(x)\mathrm{P}^{-1}=\eta^*\rho(x)\int
e^{-ip\cdot\mathcal{P}x}\rho\l{L_{\mathcal{P}p}}\r
u(k)a_p \, d\Gamma_p\quad,\\
&\mathrm{P}{\psi^-}^c(x)\mathrm{P}^{-1}=\eta^c\rho(x)\int
e^{ip\cdot\mathcal{P}x}\rho\l{L_{\mathcal{P}p}}\r
v(k){a^c}^{\dag}_p \, d\Gamma_p\quad.
\ea
\ee
We have suppressed the indices $l$ and $\sigma$ (think of the above formulas in terms of matrix multiplication). Also the change of integration variables was performed $p\rightarrow\mathcal{P}p$ and the invariance of the measure was employed.
The transformation formula may acquire a simple form when expressed in terms of causal fields if there exists a matrix 
$\stackrel{\rho}{\mathcal{P}}$, acting in the linear space of representation $\rho$, such that
\be\label{parwar1}
\ba
&\rho\l{L_{\mathcal{P}p}}\r=\stackrel{\rho}{\mathcal{P}}\rho(L_p)\stackrel{\rho}{\mathcal{P}}\quad,\\
&\stackrel{\rho}{\mathcal{P}}u(k)=b_uu(k)\quad,\quad \stackrel{\rho}{\mathcal{P}}v(k)=b_vv(k)\quad,\quad b_u,b_v\in\mathbb{C}\quad.
\ea
\ee
Then (\ref{parpsi}) is reduced to
\be\label{parwar2}
\ba
&\mathrm{P}\psi^+(x)\mathrm{P}^{-1}=\eta^*b_u\rho(x)\stackrel{\rho}{\mathcal{P}}\rho^{-1}(\mathcal{P}x)
\ \psi^+(\mathcal{P}x)\quad,\\
&\mathrm{P}{\psi^-}^c(x)\mathrm{P}^{-1}=\eta^cb_v\rho(x)\stackrel{\rho}{\mathcal{P}}\rho^{-1}(\mathcal{P}x)
\ {\psi^-}^c(\mathcal{P}x)\quad.
\ea
\ee

Let us now consider the examples. The only one--dimensional representation of the Poincar{\'e} group is the trivial one, therefore there is no need to consider scalar field. For the vector field the relevant representation of the Lorentz group is its fundamental representation in $\mathbb{R}^4$, $\rho(\Lambda){^a}_b=\Lambda{^a}_b$, which cannot be extended to the faithful representation of the whole Poincar{\'e} group in $\mathbb{R}^4$. The situation is much different for the Dirac field and the Poincar{\'e}--vector field. We shall consider these two cases separately.

\subsection{The Dirac field}

The standard spinor representation of the Lorentz group is given by the generators 
\be\label{JJ}
\mathbb{J}^{ab}=-\frac{i}{4}[\gamma^a,\gamma^b]_{-}\quad,
\ee
 where $\gamma^a$ are the Dirac matrices satisfying $[\gamma^a,\gamma^b]_{+}=2\eta^{ab}{\bf 1}$,
for which we shall choose a convenient representation
\be\label{dir}
\ba
\gamma^a=
\l
\begin{array}{cc}
0 & \sigma^a \\
\bar{\sigma}^a & 0 \\
\end{array}
\r, \quad
\sigma^0=\bar{\sigma}^0={\bf 1},\quad \bar{\sigma}^i=-\sigma^i,
\ea
\ee
where $\sigma^i$ are Pauli matrices. 
The representation admits a unique extension to the faithful representation of the Poincar{\'e} group on $\mathbb{C}^4$ provided by the generators of translations
\be\label{PP}
\mathbb{P}^a=\alpha \gamma^a(1+s\gamma^5)\quad,
\ee
where $\gamma^5=-i\gamma^0\gamma^1\gamma^2\gamma^3$, $s=\pm 1$ and $\alpha$ is a parameter of dimension of mass in natural units $c=\hbar=1$. We shall restrict ourselves to real values of $\alpha$ for which the representation $\rho$ satisfies a pseudo--unitarity condition $\rho^{\dag}(\Lambda,b)=\gamma^0\rho^{-1}(\Lambda,b)\gamma^0$.
The conditions (\ref{pchobr}), which can be imposed on the amplitudes for $j=1/2$ only because of Shur's lemma, lead to the following form 
of the amplitudes for standard momentum
\be\label{ukvk}
u(k)=
\l
\begin{array}{cccc}
c_+ & 0 \\
0 & c_+ \\
c_- & 0 \\
0 & c_-
\end{array}
\r\quad, \quad
v(k)=
\l
\begin{array}{cccc}
0 & -d_+ \\
d_+ & 0 \\
0 & -d_- \\
d_- & 0
\end{array}
\r \quad,\quad
c_+, c_-, d_+, d_- \in \mathbb{C}.
\ee
One can then calculate $u(p)$ and $v(p)$ using
\be\label{DLp}
\rho\l{L_p}\r=\rho^{\dag}\l{L_p}\r=\frac{m+p_a\gamma^a\gamma^0}{\sqrt{2m\l{p^0+m}\r}}\quad.
\ee
It can now be readily proved that the causality condition (\ref{przy}) will be fulfilled if and only if
\be\label{warprzycz}
c_+c_-^*=\pm d_+d_-^* \quad ,\quad
|c_+|^2=\mp |d_+|^2\quad,\quad
|c_-|^2=\mp |d_-|^2 \quad.
\ee
The last two equations can be satisfied only with the lower sign. It follows that the modified Dirac field necessarily describes fermions of spin $\frac{1}{2}$, just like the standard one. 

Let us finally investigate whether the theory can be made manifestly parity invariant. 
The relations (\ref{parwar1}) are valid for $\stackrel{\rho}{\mathcal{P}}=\gamma^0$. Since $\gamma^0\gamma^0={\mathbf 1}_4$, it is necessary that $b_u,b_v=\pm 1$.
It then follows from (\ref{parwar1}), (\ref{ukvk}) and (\ref{warprzycz}) that $b_v=-b_u$, $|c_-|=|c_+|=|d_-|=|d_+|=1$, $c_-=b_uc_+, d_-=-b_ud_+$. If $\psi$ is to have well established transformation properties with respect to parity, it is necessary that $\eta^c=-\eta^*$. Finally, using the possibility of changing relative phase of annihilation and creation operators and the possibility of replacing $\psi$ by $\gamma^5\psi$, we can cast the amplitudes into the standard form
\be\label{ukvk1}
u(k)=
\l
\begin{array}{cccc}
1 & 0 \\
0 & 1 \\
1 & 0 \\
0 & 1
\end{array}
\r, \quad
v(k)=
\l
\begin{array}{cccc}
0 & -1 \\
1 & 0 \\
0 & 1 \\
-1 & 0
\end{array}
\r\quad.
\ee
Note that the action of parity transformation on the field $\psi(x)$ explicitly depends on $x$, 
${\mathrm P}\psi(x){\mathrm P}^{-1}=\eta^*\rho(x)\gamma^0\rho^{-1}(\mathcal{P}x)\psi(\mathcal{P}x)$.
This dependence would not be present if the generators $\mathbb{P}$
satisfied the appropriate commutation relations with $\gamma^0$
\be\label{pbbcom}
\gamma^0\mathbb{P}^a\gamma^0={\mathcal{P}_b}^a\mathbb{P}^b
\ee
(compare (\ref{parcom})). However, the only possible nontrivial generators for the Dirac field (\ref{PP}) do not possess this property.

Since the field is of the form 
\be\label{rel}
\psi(x)=\rho(x)\tilde{\psi}(x)\quad,
\ee
where $\tilde{\psi}(x)$ possesses all the properties of the standard Dirac field, the modified field $\psi$ should satisfy the equation derived from (\ref{rel}) under the assumption that $\tilde{\psi}$ obeys the usual Dirac equation. Explicitly,
\be\label{direq}
\ba
&\l{i\gamma^a\partial_a-m}\r\tilde{\psi}(x)=0\quad\Rightarrow\quad 
\left[{\tilde{\gamma^a}(x)\l{i\partial_a+\mathbb{P}_a}\r-m}\right]\psi(x)=0\quad,\\
&\tilde{\gamma^a}(x):=\rho(x)\gamma^a\rho^{-1}(x)\quad.
\ea
\ee
This equation can be derived from the Lagrangian density
\be\label{modL0}
\mathcal{L}_0=\ov{\psi}\left[{\tilde{\gamma^a}(x)\l{i\partial_a+\mathbb{P}_a}\r-m}\right]\psi\quad,
\ee
or the Lagrangian four--form
\be
\mathfrak{L}_0=-i\l{\star dx_a}\r\wedge \ov{\psi}\tilde{\gamma^a}d\psi-\ov{\psi}\l{m-\tilde{\gamma}^a\mathbb{P}_a}\r\psi\,d^4x\quad,
\ee
where $\ov{\psi}=\psi^{\dag}\gamma^0$ is the Dirac conjugation, $\star$ is the Hodge star of Minkowski metric, i.e. $\star dx_a=\frac{1}{6}\epsilon_{abcd}dx^b\wedge dx^c\wedge dx^d$, where $\epsilon_{abcd}$ is the totally anti--symmetric symbol with $\epsilon_{0123}=1$, and $d^4x=dx^0\wedge dx^1\wedge dx^2\wedge dx^3$ is the volume form of Minkowski metric. This four--form is clearly Poincar{\'e} invariant under the global action of the relevant representation,
\be
\psi\rightarrow \rho(\Lambda,b)\psi \quad \Rightarrow\quad \ov{\psi}\rightarrow\ov{\psi}\rho^{-1}(\Lambda,b)\quad,\quad 
\tilde{\gamma}^a \rightarrow {\Lambda^a}_b\rho(\Lambda,b)\tilde{\gamma}^b\rho^{-1}(\Lambda,b)\quad,\quad dx^a\rightarrow{\Lambda^a}_bdx^b\quad 
\ee
The transformation formula for $\tilde{\gamma^a}$ follows from 
\be
\tilde{\gamma}^a(\Lambda x+b)\,=\,\rho(\Lambda x+b)\gamma^a\rho^{-1}(\Lambda x+b)\,=\,\rho(\Lambda,b)\rho(x)\rho^{-1}(\Lambda)\gamma^a\rho(\Lambda)\rho^{-1}(x)\rho^{-1}(\Lambda,b)\,=\,{\Lambda^a}_b\rho(\Lambda,b)\tilde{\gamma}^b(x)\rho^{-1}(\Lambda,b)\quad.
\ee

The fact that the new field is related via (\ref{rel}) to the standard Dirac field seems to suggest that all the physical properties of $\psi$ will be indistinguishable from those of $\tilde{\psi}$. This supposition is further supported by observation that all the Noether currents of physical importance will express in exactly the same way in terms of annihilation and creation operators when calculated for the field $\tilde{\psi}$ and $\psi$ (see the Appendix \ref{A2} for the proof). However, considering the generalized field that transforms under the faithful representation of the Poincar{\'e} group has interesting consequences for PGT. We shall discuss them in Section \ref{section2}.

\subsection{The Poincar{\'e}--vector field}\label{Poinv}

Let us now consider a faithful representation of the Poincar{\'e} group in $\mathbb{R}^5$ defined by
\be
\rho(\Lambda,b)=
\l
\begin{array}{cc}
\Lambda & \alpha b \\
0 & 1 
\end{array}
\r\quad,\quad
\Lambda\in SO(1,3)\quad,\quad b\in\mathbb{R}^4\quad,\quad \alpha\in\mathbb{R}\quad.
\ee 
The parameter $\alpha$ corresponds to the possibility of rescaling of $\mathbb{P}$ and thus is analogues to $\alpha$ that was introduced for the Dirac field.
The representation of the group of rotations that is contained in $\rho$ is a simple sum of two trivial representations and the fundamental one. The corresponding matrix acquires a block--diagonal form
\be
\rho(R)=
\l
\begin{array}{ccc}
1 & 0 & 0 \\
0 & R & 0 \\
0 & 0 & 1 
\end{array}
\r\quad,\quad
R\in SO(3)\quad,
\ee 
where zeros are zero--matrices of appropriate shapes and dimensions. Using the notation introduced in (\ref{uivi}) one can conclude that the last two equations of (\ref{pchobr}) can be solved either for $j=1,\, u_1=u_3=v_1=v_3=0$ or for 
$j=0,\,u_2=v_2=0$. 

\vskip 0.1 in
\noindent{\bf The $j=1$ case} 

This case is not really interesting, since the amplitude is then of the form
\be
u(p)=\rho(L_p)u(k)=
\l
\begin{array}{cc}
L_p & 0 \\
0   & 1  
\end{array}
\r
\l
\begin{array}{cc}
\tilde{u}(k) \\
0  
\end{array}
\r=
\l
\begin{array}{cc}
\tilde{u}(p) \\
0  
\end{array}
\r
\quad,
\ee 
where $\tilde{u}$ is the amplitude for the standard vector field. From (\ref{uxvx}) it then follows that the total $x$--dependent amplitude is of the form 
\be
u(x,p)=e^{-ip\cdot x}\rho(x)u(p)=e^{-ip\cdot x}
\l
\begin{array}{cc}
{\bf 1}_4 & \alpha x \\
0   & 1  
\end{array}
\r
\l
\begin{array}{cc}
\tilde{u}(p) \\
0  
\end{array}
\r=e^{-ip\cdot x}u(p)
\quad.
\ee
Similar result holds for $v$. The creation and annihilation fields are thus Fourier transforms of the standard momentum--dependent amplitudes for vector field of spin 1, with the unimportant row of zeros added.

\vskip 0.1 in
\noindent{\bf The j=0 case}

Since the amplitudes are of the form
\be
u(k)=
\l
\begin{array}{cc}
\frac{c_0}{m}\, k \\
c_4  
\end{array}
\r\quad,\quad
v(k)=
\l
\begin{array}{cc}
\frac{d_0}{m}\, k \\
d_4
\end{array}
\r
\quad,\quad c_0,c_4,d_0,d_4\in\mathbb{C}\quad,\quad 
\ee
where $k=(m,0,0,0)$ is the standard momentum, it follows that
\be
u(x,p)=e^{-ip\cdot x}\rho(L_p,x)u(k)=e^{-ip\cdot x}
\l
\begin{array}{cc}
\frac{c_0}{m}\, p+c_4\alpha x \\
c_4  
\end{array}
\r\quad,\quad
v(x,p)=e^{ip\cdot x}\rho(L_p,x)v(k)=e^{ip\cdot x}
\l
\begin{array}{cc}
\frac{d_0}{m}\, p+d_4\alpha x \\
d_4  
\end{array}
\r\quad
\ee
(think of $x$ and $p$ as column matrices whose entries are the components of four--momentum and Minkowskian coordinates, respectively).

Note that parity invariance does not limit the freedom of choice of the parameters at all. The relations (\ref{parwar1}) are satisfied for
\be
\stackrel{\rho}{\mathcal{P}}=
\l
\begin{array}{cc}
\mathcal{P} & 0 \\
0 & 1
\end{array}
\r\quad,\quad b_u=b_v=1\quad.
\ee
What is more, the parity transformation acts on fields in an $x$--independent way, since 
$\rho(x)\stackrel{\rho}{\mathcal{P}}\rho^{-1}(\mathcal{P}x)=\stackrel{\rho}{\mathcal{P}}$. 

The causality condition (\ref{warprzycz}) is satisfied if and only if
\be\label{caus}
\ba
&|c_0|^2(\partial_a\partial_b\triangle)(x-x')\mp |d_0|^2(\partial_a\partial_b\triangle)(x'-x)=0\quad,\\
&c_0c_4^*(\partial_a\triangle)(x-x')\mp d_0d_4^*(\partial_a\triangle)(x'-x)=0\quad,\\
&c_0^*c_4(\partial_a\triangle)(x-x')\mp d_0^*d_4(\partial_a\triangle)(x'-x)=0\quad,\\
&|c_4|^2\triangle(x-x')\mp |d_4|^2\triangle(x'-x)=0\quad
\ea
\ee
for space--like $x-x'$, where $\triangle(x):=\int e^{-ip\cdot x}d\Gamma_p$. The function $\triangle$ is even for space--like $x$, hence its derivative is odd and the second derivative is again even. Hence, the condition reduces to
\be
|c_0|^2\mp |d_0|^2=0\quad,\quad |c_4|^2\mp |d_4|^2=0\quad,\quad c_0c_4^*\mp d_0d_4^*=0\quad,\quad
\ee
that can be satisfied with the upper sign only. Hence, the particles under investigation are bosons. Adjusting the relative phase of the annihilation and creation operators and rescaling globally the field $\psi=\psi^++\psi^{-c}$ it is possible to achieve $c_4=d_4=1$. After this is done, the phases and scaling are fixed, so one cannot perform the same operation on $c_0$ and $d_0$. However, from (\ref{caus}) it now follows that $d_0=-c_0$. It seems that there are no known physical principles that could be used to limit the remaining freedom of the parameters $c_0$ and $\alpha$. It is easy to verify that the causal field thus constructed is equal to
\be\label{PVF}
\ba
&\psi(x)=\int \l{u(x,p)a_p+v(x,p)a^{c\dag}_p}\r\,d\Gamma_p=\rho(x)\tilde{\psi}(x)=
\l
\begin{array}{cc}
\Phi(x)+\alpha x\phi(x) \\
\phi(x)
\end{array}
\r\quad,\\
&\phi(x)=\int\l{e^{-ip\cdot x}a_p+e^{ip\cdot x}a^{\dag}_p}\r\,d\Gamma_p\quad,\quad 
\Phi^a(x)=\frac{ic_0}{m}\l{\partial^a\phi}\r(x)\quad,
\ea
\ee
where
\be\label{tPVF}
\tilde{\psi}=
\l
\begin{array}{cc}
\Phi \\
\phi
\end{array}
\r
\ee
satisfies the Klein--Gordon equation $\l{\square+m^2}\r\tilde{\psi}=0$. Similarly to the Dirac field case, note that $m$ is just the parameter that determines the mass shall ($p\cdot p=m^2$), and hence the invariant measure $d\Gamma_p$, and has nothing to do with $\alpha$, the latter being related to the way of embedding the Poincar{\'e} group 
in $End(\mathbb{R}^5)$. 

The simplest way of constructing Lagrangian for the field $\tilde{\psi}$ is to take a sum of Lagrangians corresponding to the $\Phi$--part and $\phi$--part. The relative weight of the two components is governed by the arbitrary parameter $c_0$, so there is no need of introducing another parameter. In a compact notation, the resulting Lagrangian density is
\be
\tilde{\mathcal{L}}=\frac{1}{2}{\partial_a\tilde{\psi}}^{T}\stackrel{\rho}{\mathcal{P}}\partial^{a}\tilde{\psi}-\frac{1}{2}m^2{\tilde{\psi}}^{T}\stackrel{\rho}{\mathcal{P}}\tilde{\psi}
\quad,
\ee 
which gives rise to the following Lagrangian density for $\psi$
\be
\ba
\mathcal{L}=
&\frac{1}{2}\left[{(\partial_a-i\mathbb{P}_a)\psi(x)}\right]^{T}\tilde{\mathcal{P}}(x)(\partial^{a}-i\mathbb{P}^a)\psi(x)-\frac{1}{2}m^2 \psi^{T}(x)\tilde{\mathcal{P}}(x)\psi(x)\quad,\\
&\tilde{\mathcal{P}}(x):=\rho^T(-x)\stackrel{\rho}{\mathcal{P}}\rho(-x)\quad.
\ea
\ee
Here the matrix $\tilde{P}$ gives rise to the explicit dependence of $\mathcal{L}$ on $x$ and plays the similar role to that of the matrices $\tilde{\gamma}^a$ in the Dirac case.

\section{The Poincar{\'e} gauge theory}\label{section2}

For any representation (\ref{Poinrep}) of the Poincar{\'e} group,
the composition law $\rho(\Lambda,b)\rho(\Lambda',b')=\rho(\Lambda\Lambda',\Lambda b'+b)$
implies the transformation properties of the generators and the commutation relations 
\be\label{transcom}
\ba
&\rho(\Lambda,b)\mathbb{P}^a\rho^{-1}(\Lambda,b)={\Lambda_c}^a\mathbb{P}^c,\\ 
&\rho(\Lambda,b)\mathbb{J}^{ab}\rho^{-1}(\Lambda,b)={\Lambda_c}^a{\Lambda_d}^b\l{\mathbb{J}^{cd}+b^c\mathbb{P}^d-b^d\mathbb{P}^c}\r , \\
&[\mathbb{P}^a,\mathbb{J}^{cd}]=-i\l{\eta^{ac}\mathbb{P}^d-\eta^{ad}\mathbb{P}^c}\r\quad, \\ 
&[\mathbb{P}^a,\mathbb{P}^b]=0, \\
&[\mathbb{J}^{ab},\mathbb{J}^{cd}]=
-i\l{\eta^{ad}\mathbb{J}^{bc}+\eta^{bc}\mathbb{J}^{ad}-\eta^{bd}\mathbb{J}^{ac}-\eta^{ac}\mathbb{J}^{bd}}\r\quad .
\ea
\ee
In order to localize the global Poincar{\'e} symmetry of special relativity and develop a gauge theory,  it is necessary to construct the covariant differential. This entails the introduction of a one--form $\mathbb{A}$ on space--time such that $i\mathbb{A}$ takes values in the representation of the Lie algebra of the group
\footnote{Since the generators of the unitary representation of the Poincar{\'e} group discussed at the beginning of the article where chosen to be Hermitian, the physicists' convention with $i$ factor in the exponential mapping is consequently followed. Hence, if $D\psi=d\psi+\mathbb{A}\psi$ is to represent the covariant derivative, then $i\mathbb{A}$ has to belong to the representation of the Lie algebra, and not $\mathbb{A}$.}
\be\label{pA}
\mathbb{A}=i\Gamma_a\mathbb{P}^a+\frac{i}{2}\omega_{ab}\mathbb{J}^{ab}\quad,
\ee
where $\omega_{ab}=-\omega_{ba}$ and $\Gamma_a$ are
one--forms on space--time, transforming as 
\be\label{gauge}
\mathbb{A}\rightarrow\mathbb{A}'=\rho(\Lambda,b)\mathbb{A}\rho^{-1}(\Lambda,b)-d\rho(\Lambda,b)\rho^{-1}(\Lambda,b)
\ee
under the local action of the group. Note that the Yang--Mills theory that we discuss is the usual one, based on linear connections. Hence, our approach is distinct from that proposed in \cite{TT1}\cite{TT2} and similar to that of \cite{GN}. From (\ref{transcom}) and (\ref{gauge}) it follows that the Yang--Mills fields
transform as
\be\label{gftrans}
\omega'=\Lambda\omega\Lambda^{-1}-d\Lambda\Lambda^{-1}, \quad
\Gamma'=\Lambda\Gamma-\omega'a-da .
\ee
Here $\omega$ is a matrix with entries
${\omega^a}_b$ and $\Gamma$ a column matrix with entries
$\Gamma^a$. 

We now come to the crucial point of PGT. Relativistic flat--space field theories are usually invariant under the mixed internal--external action of the Poincar{\'e} group, i.e. the one which acts on both fields and the Minkowskian coordinates, where the latter action is of the form $x\rightarrow x'=\Lambda x+b$. We would not have to bother with this conceptual dichotomy if we acknowledged the Minkowskian coordinates as fields. Indeed, for a given inertial observer the coordinates that they measure are just functions on space--time. We shall denote these fields by $y$, whereas the letter $x$ will refer to arbitrary coordinates from now on. Hence, the field $y$ and its differential transform under the global Poincar{\'e} transformations as
\be\label{ytrans}
y\rightarrow \Lambda y+b\quad,\quad dy\rightarrow \Lambda y\quad .
\ee
How to construct its covariant differential? No homogeneous expression of the form $Dy=dy+Ay$ can yield the correct transformation properties under the local action of the full Poincar{\'e} group. In particular, contrary to the statement of \cite{GN}, the ``covariant derivative'' $Dy=dy+\omega y$ would not transform as $y$ itself. To see this, consider a local pure translation $y\rightarrow y+b$ for which it follows that $Dy\rightarrow Dy+db+\omega b\not= Dy+b$. However, we are seeking for the covariant differential that transforms as $dy$, rather than $y$ (i.e. homogeneously). It follows from (\ref{gftrans}) and (\ref{ytrans}) that 
\be\label{Dy}
\mathcal{D}y=dy+\omega y+\Gamma
\ee
works well, i.e. it transforms as
\be\label{Dytrans}
\mathcal{D}y\rightarrow \Lambda\mathcal{D}y
\ee
under local Poincar{\'e} transformation $(\Lambda,b)$.

We have considered two examples of fields in Minkowski space that transform under faithful representations of the Poincar{\'e} group. We shall now consider the inclusion of gravity for the theory of modified Dirac field (for the Poincar{\'e}--vector field the procedure works similarly). According to the notation of this section, the Lagrangian density generating the appropriate field equation (\ref{direq}) for the modified Dirac field $\psi$ is 
\be\label{Lkl}
\mathcal{L}_{\kappa\lambda}=\ov{\psi}\left[{\tilde{\gamma^a}(y)\l{i\partial_a+\mathbb{P}_a}\r-m}\right]\psi+\kappa\,\partial_aJ^a_{(V)}+\lambda\,\partial_aJ^a_{(A)}\quad,
\ee
where $\kappa,\lambda\in\mathbb{C}$ are constants and $J^a_{(V)}=\ov{\psi}\tilde{\gamma}^a\psi=\ov{\tilde{\psi}}\gamma^a\tilde{\psi}$ 
and $J^a_{(A)}=\ov{\psi}\tilde{\gamma}^a\tilde{\gamma}^5\psi=\ov{\tilde{\psi}}\gamma^a\gamma^5\tilde{\psi}$ are the Dirac vector and axial currents. All the Lagrangian densities corresponding to different values of $\kappa$ and $\lambda$ generate the correct field equations, since adding divergence of a vector field to the Lagrangian density results in a surface term in the action that vanishes when varieted with appropriate boundary conditions. In spite of this, the theories with gravity obtained for different choices of the parameters via the standard minimal coupling procedure would not be equivalent \cite{Kazm1}. We shall comment on this more at the end of this section. The corrected unambiguous coupling procedure will be reviewed. Now we shall use minimal coupling and choose $\kappa=-\frac{i}{2}, \lambda=0$, in order to clarify the relation of the Lagrangian to the one proposed in \cite{GN}. The resulting Lagrangian density is
\be\label{LR}
\mathcal{L}_R=\frac{i}{2}\l{\partial_a\ov{\psi}\tilde{\gamma}^a\psi-\ov{\psi}\tilde{\gamma}^a\partial_a\psi}\r-(m-4\alpha)\ov{\psi}\psi\quad.
\ee
To derive this result, use $[\tilde{\gamma}^a,\mathbb{P}_a]_+=[\gamma^a,\mathbb{P}_a]_+=8\alpha{\bf 1}$. Note that this form suggests that the mass of the field is $m-4\alpha$. However, when rewritten in terms of $\tilde{\psi}=\rho^{-1}(y)\psi$, the Lagrangian reads 
$\mathcal{L}_R=\frac{i}{2}\l{\partial_a\ov{\tilde{\psi}}\gamma^a\psi-\ov{\tilde{\psi}}\gamma^a\partial_a\psi}\r-m\ov{\tilde{\psi}}\tilde{\psi}$. All the Noether currents will look like those of the standard Dirac field of mass $m$, when written in terms of annihilation and creations operators. Therefore, our claim is that the real physical mass of the field $\psi$ is $m$. One can also give another argument: (\ref{psitr}) implies that
\be
e^{ib\cdot P}\psi(y) e^{-ib\cdot P}=e^{-ib\cdot\mathbb{P}}\psi(y+b)\quad
\ee
for arbitrary $b$. Expanding this equality in $b$ up to quadratic terms we get
\be
[P^a,[P_a,\psi]_-]_-=-\square\psi+2i\mathbb{P}^a \partial_a\psi\quad,
\ee
where $\square:=\partial_a\partial^a$ is the Dalambert operator. On the other hand, the field equation implies that 
\be
0=\left[{\tilde{\gamma}^a\l{i\partial_a+\mathbb{P}_a}\r+m}\right]\left[{\tilde{\gamma}^a\l{i\partial_a+\mathbb{P}_a}\r-m}\right]\psi=
-\square\psi+2i\mathbb{P}^a\partial_a\psi-m^2\psi\quad.
\ee
It follows that $[P^a,[P_a,\psi]_-]_-=m^2\psi$, so $m^2$ is the eigen--value of the Casimir operator $P\cdot P$, hence $m$, and not $m-4\alpha$, plays the role of the mass.
Note that the value of $m$ can be safely set to zero. The field would then be massless, in spite of the apparent ``mass'' of $-4\alpha$ appearing in the Lagrangian when written in the form (\ref{LR}). No inconsistencies occur in the formalism in the $m=0$ case. We do not see the arguments supporting the statement of \cite{GN} according to which only massive fermions can carry the representation of the full Poincar{\'e} group, although the discussion of Section \ref{QFT} certainly ought to be modified to account for the massless case appropriately.

In order to turn the gravitation on, it is convenient to use the Lagrangian four--form, rather than Lagrangian density
\be
\mathfrak{L}_R\l{\psi,\ov{\psi},y,d\psi,d\ov{\psi},dy}\r=-\frac{i}{2}\l{\star dy_a}\r\wedge 
\left[{d\ov{\psi}\,\tilde{\gamma}^a(y)\,\psi-\ov{\psi}\,\tilde{\gamma}^a(y)\,d\psi}\right]-(m-4\alpha)\ov{\psi}\psi\,dy^0\wedge dy^1\wedge dy^2 \wedge dy^3\,.
\ee
The minimal coupling $d\psi\rightarrow D\psi=d\psi+\mathbb{A}\psi$, $dy\rightarrow \mathcal{D}y$, with $\mathbb{A}$ being given by (\ref{pA}) and $\mathcal{D}y$ by (\ref{Dy}), leads to the following four--form that includes gravitational interaction
\be\label{LRgr}
\tilde{\mathfrak{L}}_R=
-\frac{i}{2}\l{\star \mathcal{D}y_a}\r\wedge 
\l{D\ov{\psi}\,\tilde{\gamma}^a(y)\,\psi-\ov{\psi}\,\tilde{\gamma}^a(y)\,D\psi}\r-(m-4\alpha)\ov{\psi}\psi\,\epsilon\,,
\ee
where $\epsilon=\mathcal{D}y^0\wedge \mathcal{D}y^1\wedge \mathcal{D}y^2 \wedge \mathcal{D}y^3$. 

Note that the covariant differential is the usual Yang--Mills one that transforms as $D\psi\rightarrow \rho(\Lambda,b)D\psi$ under the action of a local Poincar{\'e} transformation $(\Lambda,b)$.
 Another form of covariant differential was employed in \cite{TT1}\cite{TT2} within the framework of nonlinear realizations of PGT, 
\be\label{nlcov}
\check{D}\check{\psi}=d\check{\psi}+\frac{i}{2}\omega_{ab}\mathbb{J}^{ab}\check{\psi}+ie^a\mathbb{P}_a\check{\psi}\quad,
\ee
where $\check{}$ refers to the objects defined in \cite{TT1}\cite{TT2} and $e^a$ are one--forms that transform as $e^a\rightarrow{\Lambda^a}_be^b$ under the action of $(\Lambda,b)$ and have geometrical interpretation of a cotetrad. Since the field $\check{\psi}$ that appears in the nonlinear framework transforms trivially under translations, such covariant derivative transforms in a reasonable way, i.e. $\check{D}\check{\psi}\rightarrow \rho(\Lambda)\check{D}\check{\psi}$ under the transformation $(\Lambda,b)$. In the case of our field $\psi$ that transforms under the faithful representation of the Poincar{\'e} group, the form (\ref{nlcov}) of a covariant derivative would not be appropriate, since the transformation properties under translations would be very inconvenient.

It is now time to recall that we aim to formulate the
theory of gravitational interaction, which ought to be bound up with
the geometry of space--time, according to Einstein's idea. The transformation formulas (\ref{Dytrans}) and (\ref{gftrans}) suggest the interpretation of $\mathcal{D}y^a$ as a cotetrad, which we shall denote by $e^a:=\mathcal{D}y^a$ from now on, and the interpretation of $\omega_{ab}$ as connection one--forms in the orthonormal frame $e^a$. Note that the anti--symmetry of $\omega_{ab}$ is equivalent to the metricity of the connection (with respect to the unique metric in which the cotetrad is orthonormal, given explicitly by $g=\eta_{ab}e^a\otimes e^b$, where $\otimes$ is the tensor product). Note that encoding the space--time metric in the cotetrad is particularly convenient if the Dirac field is present, which was noticed already by Weil \cite{Weil}. Hence, the space--time acquired the structure of a Riemann--Cartan manifold $\mathcal{M}(e,\omega)$, i.e. a manifold with the metric and
 the metric--compatible connection defined on it.

The Lagrangian four--form (\ref{LRgr}) has the form of that of reference \cite{GN} for $\alpha=m$, up to the ``mass term'', which was not written down there. 
To see this, use the Backer--Campbell--Hausdorf formula and show that
\be\label{tildeg}
\tilde{\gamma}^a(y)=e^{iy\cdot\mathbb{P}}\gamma^ae^{-iy\cdot\mathbb{P}}=\gamma^a-i\alpha y^b
\left[{\gamma^a\gamma_b({\bf 1}+s\gamma^5)-\gamma_b\gamma^a({\bf 1}-s\gamma^5)}\right]+
4\alpha^2\l{y^a\gamma_b y^b-\frac{1}{2}y^by_b\gamma^a}\r 
({\bf 1}+s\gamma^5)\quad .
\ee
That the Lagrangian (\ref{LRgr}) is invariant under local Poincar{\'e} transformations is clear from the construction and can be verified explicitly, although the expended form of $\tilde{\gamma}(y)$ in (\ref{tildeg}) is vary inconvenient for this calculation. This is probably why the issue of Poincar{\'e} invariance of the Lagrangian presented in \cite{GN} was questioned in \cite{Lec1}. Hopefully, the discussion presented here will dispel these doubts. The relevant transformation formula for $\gamma^a$ (without $^{\tilde{}}$) is obviously the trivial one, i.e. the matrices $\gamma^a$ do not transform.

If we had applied the minimal coupling to the general Lagrangian (\ref{Lkl}), we would have obtained a two--complex--parameter family of non--equivalent theories in Riemann--Cartan space, not all of them being consistent. If the parameters were chosen such that the Lagrangian was real, the theories obtained would be consistent but still non--equivalent. The resulting ambiguity has meaningful physical consequences \cite{Kazm1,Kazm2}. Luckily, it can be avoided by using the modified covariant derivative 
\be\label{calD}
\mathcal{D}\psi=D\psi+\frac{1}{2}T_ae^a\psi\quad,
\ee
instead of $D\psi$. Here $T_a:={T^b}_{ab}$ is the torsion trace vector field. The torsion components can be defined via the Cartan structure equation
$\frac{1}{2}{T^a}_{bc}e^b\wedge e^c=de^a+{\omega^a}_b\wedge e^b$. See \cite{Kazm4} for the discussion of this solution and its uniqueness under certain natural assumptions. 
Note that the modified covariant derivative (\ref{calD}) is in fact different form the one discussed in \cite{Kazm4} (although it looks formally the same), since $D\psi$ contains now both the connection and the translational gauge fields. The usage of (\ref{calD}) in coupling procedure when applied to (\ref{Lkl}) leads to the two--complex--parameter family of equivalent Lagrangians in Riemann--Cartan space. The corresponding actions differ by topological terms depending on $\kappa$ and $\lambda$. The Lagrangian four--form $\tilde{\mathcal{L}}_R$ belongs to this family.

\section{Conclusions and open problems}\label{conc}

Although the fields that transform under a faithful representation of the full Poincar{\'e} group can be consistently constructed within the framework of QFT, they are necessarily of the form $\psi(x)=\rho({\mathbf 1},x)\tilde{\psi}(x)$, where $\rho(\Lambda,b)$ is the defining representation of the Poincar{\'e} group and $\tilde{\psi}$ transforms under the representation of the Lorentz group $\rho(\Lambda,0)$. If $\rho(\Lambda,0)$ is an example of a representation of the Lorentz group that is of interest in standard QFT, then the theory of a generalized field $\psi$ is essentially physically equivalent to the theory of a field $\tilde{\psi}$. This was demonstrated in details on the example of Dirac field. If, however, $\rho(\Lambda,0)$ is a simple sum of irreducible representations that are not mixed by discrete symmetries, then there is no reason for considering fields transforming under $\rho(\Lambda,0)$ in standard QFT. One is rather interested in the particular constituent irreducible representations and the fields defined by them. However, there are representations of the Poincar{\'e} group that are not completely reducible, but contain a completely reducible representation $\rho(\Lambda,0)$ of a Lorentz group. The parts of the field that transform independently under $\rho(\Lambda,0)$ are then mixed by translations. Attributing physical significance to such a representation makes the corresponding representation $\rho(\Lambda,0)$ important. The resulting theory is certainly equivalent to a theory of some combination of standard fields, but the way of combining them together is determined by  $\rho$. 
An example of a situation of this kind is provided by the Poincar{\'e}--vector representation discussed in Section \ref{QFT} that is neither irreducible nor completely reducible -- see e.g.\cite{Hall} for the definitions). 

Hence, classifying fields by the representations of the Poincar{\'e} group may lead to merging some known Lorentz--transforming fields into larger entities of well established value of spin and mass. The Lagrangians describing such generalized fields will necessarily depend explicitly on the space--time coordinates. 
The possibility of introducing such generalized fields leads to the conclusion that the translational gauge fields can appear in the covariant derivative of matter fields, after gravity is included along the lines of PGT. Such a procedure, when applied to the Dirac field, leads to the Lagrangian which closely resembles the one introduced in \cite{GN} (the only difference concerns the mass term). The discussion presented in Section \ref{section2} seems to show that the interpretation of our parameter $\alpha$ as the mass of the field, which interpretation was suggested in \cite{GN} and \cite{TT1}, is not really supported, at least on the ground of standard Yang--Mills theory based on linear connection (i.e. the theory considered in \cite{GN} and in our paper). Therefore, the issue of physical interpretation of this parameter, as well as the additional parameter $c_0$ that is involved in the discussion of Poincar{\'e}--vector field, remains open. It would be interesting to inspect the relation of our Poincar{\'e}--vector field of spin $0$ to the ``scalar'' field discussed in \cite{GN}. It could be also extremely interesting to discuss other finite--dimensional representations of the Poincar{\'e} group that are not completely reducible and couple gravity to fields thus constructed. Although the resulting theories will then be equivalent to those obtained by merging some standard fields with the Poincar{\'e} coordinates of PGT, the way of merging will be uniquely prescribed by the underlying representation of the Poincar{\'e} group and the fact that translations mix the particular components may in principle lead to interesting physical ramifications.

\section*{Acknowledgements}
I wish to thank prof. J. Lewandowski for helpful comments.
This work was partially supported by 
the Foundation for Polish Science, grant ''Master''.

\section{Appendix: Notation and conventions}\label{A1}

Throughout the paper $a,b,\dots$ are orthonormal tetrad indices and $\mu,\nu,\dots$ correspond to
a holonomic frame. For inertial frame of flat Minkowski space,
which is both holonomic and orthonormal, we use $a,b,\dots\in\{0,1,2,3\}$ for the whole space--time and $i,j,\dots\in\{1,2,3\}$ for the spatial section.
The metric components in an orthonormal tetrad basis $\tilde{e}_a$ are
 $g\l{\tilde{e}_a,\tilde{e}_b}\r=(\eta_{ab})=diag(1,-1,-1,-1)$ and the dual basis of one--form fields (the cotetrad) is denoted by $e^a$ 
(hence, $e^a(\tilde{e}_b)={\delta^a}_b$). Lorentz
 indices are shifted by $\eta_{ab}$. $\epsilon=e^0\wedge e^1\wedge
e^2\wedge e^3$ denotes the canonical
volume four--form  whose components in orthonormal tetrad basis obey $\epsilon_{0123}=-\epsilon^{0123}=1$.
The Hodge star action on external products of orthonormal cotetrad
one--forms is given by
\beq
\star e_a=\frac{1}{3!}\epsilon_{abcd}e^b\w e^c\w e^d \ , \quad 
\star \l{e_a\w e_b}\r=\frac{1}{2!}\epsilon_{abcd}e^c\w e^d \ , \quad 
\star \l{e_a\w e_b\w e_c}\r=\epsilon_{abcd}e^d \ ,
\eeq
which by linearity determines the action of $\star$ on any differential
form.

\section{Appendix: Noether theorem}\label{A2}

Let  
\be
S[\Phi^A]=\int\mathcal{L}\l{\Phi^A,\partial_{\mu}\Phi^A}\r\d^4 x
\ee
represent  the action of a field theory on a smooth manifold $\mathcal{M}$ (which is not necessarily the Minkowski space). Here $x^{\mu}$ are arbitrary coordinates and hence $\mathcal{L}$ is a scalar density. Let $\mathcal{T}$ be the target
space in which the collection of fields $\Phi$ take its values. 
Consider a Lie group $\mathcal{G}$ that acts on $\mathcal{T}$ as a group of transformations. Let 
\be\label{symtr}
\ba
\Phi^A\longrightarrow \Phi'^A=\Phi^{A}+\delta \Phi^{A} 
\ea
\ee
represent the infinitesimal form of the action of $\mathcal{G}$ on
$\mathcal{T}$. The transformations are called 
{\it symmetry transformations} if they do not change the
action, up to possibly surface terms (and thus leave the form of field
equations invariant). This is equivalent to
\be\label{symcond}
\frac{\partial\mathcal{L}}{\partial\Phi^A}\delta\Phi^A+ 
\frac{\partial \mathcal{L}}{\partial(\partial_{\mu}\Phi^A)}
\partial_{\mu}\delta\Phi^A=\partial_{\mu}W^{\mu}\quad ,
\ee
where $W^{\mu}$ is a vector density. This can be further expressed as
\beq
\partial_{\mu}j^{\mu}=
\l{\partial_{\mu} 
\frac{\partial \mathcal{L}}{\partial(\partial_{\mu}\Phi^A)}-\frac{\partial\mathcal{L}}{\partial\Phi^A}}\r
\delta\Phi^A \quad ,
\eeq
where 
\be\label{j}
j^{\mu}=\frac{\partial\mathcal{L}}{\partial(\partial_{\mu}\Phi^A)}\delta\Phi^A-W^{\mu} \quad 
\ee
is a Noether current associated to the symmetry transformation (\ref{symtr}), which
is clearly conserved, i.e. $\partial_{\mu}j^{\mu}=0$, if the
Euler--Lagrange equations for fields are satisfied.

An interesting class of transformations in Minkowski space is constituted by these transformations that act on both the fields and the Minkowskian coordinates $y^a$. The discussion of Noether theorem presented above applies to this case if the coordinates $y^a(x)$ are interpreted as additional fields on space--time. As explained in Section 
\ref{section2}, this way of viewing Minkowskian coordinates is very convenient in PGT. The set of all fields $\Phi^A$ consists then of matter fields $\phi^m$ and the Poincar{\'e} coordinates $y^a$. The Lagrangian density is of the form
\be
\mathcal{L}=\pounds \mathrm{det} J\quad,
\ee 
where $\pounds$ is a scalar part of $\mathcal{L}$ (which coincides with $\mathcal{L}$ in Minkowskian coordinates) and $J^a_{\mu}:=\partial_{\mu}y^a$ is the Jacobi matrix. In this case, the conserved current (\ref{j}) can be rewritten in a more convenient form as
\be\label{rmj}
\mathrm{j}^a:=\frac{1}{\mathrm{det} J}J^a_{\mu}j^{\mu}=\frac{\partial\pounds}{\partial(\partial_a\phi^m)}\delta\phi^m-
\left[{\frac{\partial\pounds}{\partial(\partial_a\phi^m)}\partial_b\phi^m-\delta^a_b\pounds}\right]\delta y^b-
\frac{1}{\mathrm{det} J}J^a_{\mu}W^{\mu} \quad .
\ee
In the calculation above the identities 
\be
\frac{\partial\pounds}{\partial(\partial_{\mu}\phi^m)}=J^{\mu}_a \frac{\partial\pounds}{\partial(\partial_a\phi^m)}\quad,\quad 
\frac{\partial\pounds}{\partial(\partial_{\mu}y^a)}=-J^{\mu}_b\partial_a\phi^m \frac{\partial\pounds}{\partial(\partial_b\phi^m)}\quad,\quad 
\frac{\partial\mathrm{det}J}{\partial(\partial_{\mu}y^a)}=\mathrm{det}JJ^{\mu}_a
\ee 
where used. Note that $\mathrm{j}$ is a vector field (not a vector density), whose components in the basis $\partial_{\mu}$ are 
$\mathrm{j}^{\mu}=J^{\mu}_a\mathrm{j}^a=\frac{1}{\mathrm{det} J}j^{\mu}$. Hence, if the Euler--Lagrange equations hold, then $0=\partial_{\mu}\l{\mathrm{det}J\,\,\mathrm{j}^{\mu}}\r=\mathrm{det}J\nabla_{\mu}\mathrm{j}^{\mu}$, where $\nabla$ is the Levi--Civita connection of the Minkowski metric. Therefore, $\nabla_{\mu}\mathrm{j}^{\mu}=\nabla_a\mathrm{j}^a=0$. In Minkowskian coordinates the covariant derivative reduces to partial derivative and hence (\ref{rmj}) is also conserved, i.e. $\partial_a\mathrm{j}^a=0$.

Let us now assume that the Lagrangian density is invariant\footnote{It means that the transformations are symmetries and the corresponding wektor density $W^{\mu}$ is equal to zero.} under the action of the group of space--time translations that acts on $y^a$ as $y^a\rightarrow y^a+\lambda b^a$, where $\lambda$ is an infinitesimal parameter of a transformation and $b$ is an element of $\mathbb{R}^4$. In conventional field theory, the matter fields do not transform under translations, so $\delta\phi^m=0$ and $\delta y^a=\lambda b^a$ imply that the canonical energy--momentum tensor in the form
\be
{t_b}^a=\frac{\partial\pounds}{\partial(\partial_a\phi^m)}\partial_b\phi^m-\delta^a_b\pounds
\ee
is conserved, $\partial_a {t_b}^a=0$. Let us now consider the modified Dirac field, with the Lagrangian given by (\ref{modL0}). Note however that the letter $x$ that appears there ought to be replaced by $y$ according to the notation we use here, since it refers to Minkowskian coordinates. Note also that $\mathcal{L}_0\equiv\pounds_0$ and that $\psi$ and $\ov{\psi}$ have to be considered as independent fields. Since $\delta\psi=i\lambda b\cdot\mathbb{P}\psi$ and $\delta\ov{\psi}=-i\lambda\ov{\psi}b\cdot\mathbb{P}$ (think of $\psi$ and $\ov{\psi}$ as a column and raw matrix respectively), it follows that
\be
\mathrm{j}^a=-\lambda b^b\l{\ov{\psi}\tilde{\gamma}^a(y)\mathbb{P}_b\psi+i\ov{\psi}\tilde{\gamma}^a(y)\partial_b\psi-\delta^a_b\mathcal{L}_0}\r
\ee
and hence the appropriate energy--momentum tensor for the field $\psi$ is
\be
{t_b}^a=\ov{\psi}\tilde{\gamma}^a(y)\mathbb{P}_b\psi+i\ov{\psi}\tilde{\gamma}^a(y)\partial_b\psi-\delta^a_b\mathcal{L}_0\quad.
\ee
The presence of the first component makes this expression look differently from the conventional energy--momentum tensor for the Dirac field. Recall, however, that $\psi=\rho(y)\tilde{\psi}$, where $\rho(y)=\exp\l{iy\cdot\mathbb{P}}\r$ and $\tilde{\psi}$ is the usual Dirac field that transforms trivially under translations. Since 
$i\ov{\psi}\tilde{\gamma}^a(y)\partial_b\psi=i\ov{\tilde{\psi}}\gamma^a\partial_b\tilde{\psi}-\ov{\tilde{\psi}}\gamma^a\mathbb{P}_b\tilde{\psi}$, it follows that ${t_b}^a$ is the standard energy--momentum tensor when expressed in terms of $\tilde{\psi}$. Similarly, the conserved current that is related to the symmetry under the change of a phase of $\psi$ is $\ov{\psi}\tilde{\gamma}^a(y)\psi=\ov{\tilde{\psi}}\gamma^a\tilde{\psi}$. Therefore, these currents will express trough the annihilation and creation operators in exactly the same way as in the standard theory.

Certainly, the interpretation of Minkowskian coordinates as fields, which is useful in PGT, is not necessary to discuss Noether theorem (see e.g. the Appendix of \cite{Kazm1} for more conventional approach).

\end{document}